\documentclass[submission,copyright,creativecommons]{eptcs}
\providecommand{\event}{PLACES 2022} 
\usepackage{breakurl}             
\usepackage{underscore}           
\usepackage{xcolor}
\usepackage{amsmath}
\usepackage{amsfonts}
\usepackage{amssymb}
\usepackage{cmll}
\usepackage{inconsolata}

\title{Replicate, Reuse, Repeat:
Capturing Non-Linear Communication via Session Types and Graded Modal Types}

\author{Daniel Marshall
\institute{University of Kent, UK}
\email{dm635@kent.ac.uk}
\and
Dominic Orchard
\institute{University of Kent \& University of Cambridge, UK}
\email{d.a.orchard@kent.ac.uk}}
\def\titlerunning{Capturing Non-Linear Communication via Session Types and Graded Modal Types}
\def\authorrunning{D. Marshall \& D. Orchard}

\newcommand{\dnote}[1]{}
\newcommand{\dmnote}[1]{}

\definecolor{coeffectColor}{HTML}{0750d0}
\definecolor{effectColor}{HTML}{d64800}

\usepackage{listings}
\lstdefinelanguage{Granule}{%
  mathescape=true,
  morecomment=[l]{--},
  moredelim=[s][\itshape]{`}{`},
  showspaces=false,
  xleftmargin=2.5em,
  commentstyle=\itshape\color{black!60},
  basicstyle=\ttfamily\footnotesize,
  flexiblecolumns=true,
  columns=[l]flexible,
  columns=fullflexible,
  keepspaces=true,
  literate=%
  {<}{\textcolor{effectColor}{<}}1
  {>}{\textcolor{effectColor}{>}}1
  {[}{\textcolor{coeffectColor}{[}}1
  {]}{\textcolor{coeffectColor}{]}}1
  {[r' : R']}{[{\textcolor{coeffectColor}{r' : R'}}]}1
  {[([}{\textcolor{coeffectColor}{[[}}2  
  {])]}{\textcolor{coeffectColor}{]]}}2
  {!a}{\textcolor{coeffectColor}{!a}}1
  {!b}{\textcolor{coeffectColor}{!b}}1
  {!Cake}{\textcolor{coeffectColor}{!Cake}}1
  {*Coffee}{\textcolor{uniqueColor}{*Coffee}}1
  {*a}{\textcolor{uniqueColor}{*a}}1
  {*FloatArray}{\textcolor{uniqueColor}{*FloatArray}}1
  {forall}{$\forall$}1
  {Inf}{$\infty$}1
  {->}{$\rightarrow$}1
  {-o}{$\multimap$}1
  {=>}{$\Rightarrow$}1
  {<-}{\textcolor{effectColor}{$\leftarrow$}}1
  {/\\}{$\sqcap$}1
  {\\/}{$\sqcup$}1
  {<=}{$\leqslant$}1
  {>=}{$\geqslant$}1
  {\\}{$\lambda$}1
  {_1}{$\mathtt{_1}$}1
  {_2}{$\mathtt{_2}$}1
  {_3}{$\mathtt{_3}$}1
  {_4}{$\mathtt{_4}$}1
  {_L}{$\mathtt{_{L}}$}1
  {_LH}{$\mathtt{_{LH}}$}1
  {_Gr}{$\mathtt{_{Gr}}$}1
  {_p}{$\mathtt{_{p}}$}1
  {_q}{$\mathtt{_{q}}$}1
  {-o}{$\multimap$}1
  {\\times}{$\times$}1
  {--BLANK}{}1,
  keywordstyle = \color{blue!70!black},
  keywords = {data, type, let, in, case, of, if, then, else, where,import, Type, Semiring, Protocol, Nat, Predicate},
  keywordstyle = [3]\color{purple!70!black},
  keywords = [3]{Send, Recv, Dual, Select, Offer, End},
  keywordstyle = [4]\bfseries\color{effectColor},
  keywords     = [4]{pure},
  numbers=left,
  numberstyle=\tiny\color{gray}
}

\lstnewenvironment{granule}
{\lstset{language=Granule}}{}

\lstnewenvironment{granuleInterface}
{\lstset{language=Granule,numbers=none}}{}

\newcommand{\granin}[1]{\lstinline[language=Granule]{#1}}

\definecolor{multiplicity}{rgb}{0,0.3,0.08}

\lstdefinelanguage{Haskell}{%
  mathescape=true,
  morecomment=[l]{--},
  comment=[l]{\{-},
  moredelim=[s][\itshape]{`}{`},
  showspaces=false,
  commentstyle=\itshape\color{black!60},
  basicstyle=\ttfamily\footnotesize,
  flexiblecolumns=true,
  columns=[l]flexible,
  columns=fullflexible,
  keepspaces=true,
  xleftmargin=2.5em,
  literate=%
   {\%r}{\textcolor{multiplicity}{\%r}}1
   {\%'Many}{\textcolor{multiplicity}{\%'Many}}1
   {\%1}{\textcolor{multiplicity}{\%1}}1,
  keywordstyle = \color{blue!40!black},
  keywords = {data, type, let, in, case, of, if, then, else, where,
  import, class, instance},
  numbers=left,
  numberstyle=\tiny\color{gray}
}

\lstnewenvironment{haskell}
{\lstset{language=Haskell}}{}

\newcommand{\haskin}[1]{\lstinline[language=Haskell]{#1}}

\lstdefinelanguage{Clean}{%
  mathescape=true,
  morecomment=[l]{//},
  comment=[l]{//},
  moredelim=[s][\itshape]{`}{`},
  showspaces=false,
  commentstyle=\itshape\color{black!40},
  basicstyle=\ttfamily\footnotesize,
  flexiblecolumns=true,
  columns=[l]flexible,
  columns=fullflexible,
  keepspaces=true,
  xleftmargin=2.5em,
  literate=%
   {*\{Real\}}{\textcolor{uniqueColor}{*\{Real\}}}1
   {*Coffee}{\textcolor{uniqueColor}{*Coffee}}1
   {*Awake}{\textcolor{uniqueColor}{*Awake}}1,
  keywordstyle = \color{blue!40!black},
  keywords = {data, type, let, in, case, of, if, then, else, where,
  import, class, instance},
  numbers=left,
  numberstyle=\tiny\color{gray}
}

\lstnewenvironment{clean}
{\lstset{language=Clean}}{}

\usepackage[most]{tcolorbox}

\makeatletter
\newtcblisting[auto counter]{haskellp}[1][]{%
  enhanced jigsaw,
  colframe=black!30,
  boxrule=0.5pt,
  toprule=0.5pt,
  top=0em,
  left=0em,
  bottom=0em,
  bottomrule=0.5pt,
  boxsep=0em,
  after skip=1em,
  before skip=1em,
  listing options={language={haskell}},
  listing only,
  colback=white,
  sharp corners,
  overlay={\node[inner sep=4pt,xshift=-2cm,fill=white] (A) at (frame.south east) {Haskell};
  },
  #1,
}
\makeatother

\newtcblisting[auto counter]{cleanp}[1][]{%
  enhanced jigsaw,
  colframe=black!30,
  boxrule=0.5pt,
  toprule=0.5pt,
  top=0em,
  left=0em,
  bottom=0em,
  bottomrule=0.5pt,
  boxsep=0em,
  after skip=1em,
  before skip=1em,
  listing options={language={clean}},
  listing only,
  colback=white,
  sharp corners,
  overlay={\node[inner sep=4pt,xshift=-2cm,fill=white] (A) at (frame.south east) {Clean};
  },
  #1,
}
\makeatother

\newtcblisting[auto counter]{cleanpill}[1][]{%
  enhanced jigsaw,
  colframe=black!30,
  boxrule=0.5pt,
  toprule=0.5pt,
  top=0em,
  left=0em,
  bottom=0em,
  bottomrule=0.5pt,
  boxsep=0em,
  after skip=1em,
  before skip=1em,
  listing options={language={clean}},
  listing only,
  colback=white,
  sharp corners,
  overlay={\node[inner sep=4pt,xshift=-2cm,fill=white] (A) at (frame.south east) {Ill-typed Clean};
  },
  #1,
}
\makeatother

\makeatletter
\newtcblisting[auto counter]{haskellf}[1][]{%
  top=0em,
  left=0em,
  bottom=0em,
  bottomrule=0.5pt,
  boxsep=0em,
  after skip=1em,
  before skip=1em,
  listing options={language={haskell}},
  listing only,
  colback=white,
  #1,
}
\makeatother

\makeatletter
\newtcblisting[auto counter]{granulep}[1][]{%
  enhanced jigsaw,
  colframe=black!30,
  boxrule=0.5pt,
  toprule=0.5pt,
  bottomrule=0.5pt,
  top=0em,
  left=0em,
  bottom=0em,
  boxsep=0em,
  after skip=1em,
  before skip=1em,
  listing options={language={granule}},
  listing only,
  colback=white,
  sharp corners,
  overlay={
           \node[inner sep=2pt,xshift=-2cm,fill=white] (B) at (frame.south east) {Granule};
  },
  #1,
}
\makeatother

\makeatletter
\newtcblisting[auto counter]{granulepill}[1][]{%
  enhanced jigsaw,
  colframe=black!30,
  boxrule=0.5pt,
  toprule=0.5pt,
  bottomrule=0.5pt,
  top=0em,
  left=0em,
  bottom=0em,
  boxsep=0em,
  after skip=1em,
  before skip=1em,
  listing options={language={granule}},
  listing only,
  colback=white,
  sharp corners,
  overlay={
           \node[inner sep=2pt,xshift=-2cm,fill=white] (B) at (frame.south east) {Ill-typed Granule};
  },
  #1,
}
\makeatother

\begin{document}
\maketitle

\begin{abstract}
  Session types provide guarantees about concurrent behaviour and can be
  understood through their correspondence with linear logic, with propositions as
  sessions and proofs as processes. However, a strictly linear setting is
  somewhat limiting, since there exist various useful patterns of communication
  that rely on non-linear behaviours. For example, shared channels provide a way
  to repeatedly spawn a process with binary communication along a fresh linear
  session-typed channel. Non-linearity can be introduced in a controlled way in
  programming through the general concept of \emph{graded modal types}, which
  are a framework encompassing various kinds of \emph{coeffect} typing
  (describing how computations make demands on their context).
  This paper shows how graded modal types can be leveraged alongside session
  types to enable various non-linear concurrency behaviours to be re-introduced
  in a precise manner in a type system with a linear basis. The ideas here are
  demonstrated using Granule, a functional programming language with linear,
  indexed, and graded modal types.
\end{abstract}

\section{Introduction}

Most programming languages ascribe a notion of \emph{type} (dynamic or static)
to data, classifying \emph{what} data we are working with---integer, string,
function, etc. \emph{Behavioural types} on the other hand capture not just
\emph{what} data is, but \emph{how} it is calculated. One such behavioural type
system for concurrency is \emph{session types}~\cite{honda}, representing the
behaviour of a process communicating over a channel via the channel's type.
Session types naturally fit into the more general substructural discipline of
\emph{linear types}~\cite{girard,wadler,walker2005substructural}.

By treating data as a \emph{resource} which must be used exactly once, linear
types can capture various kinds of stateful protocols of interaction.
Session types are inherently substructural:
channels cannot in general be arbitrarily duplicated or
discarded and must be used according to a sequence of operations (the
protocol). This idea has allowed logical foundations to be developed for
session types based on Curry-Howard correspondences with both intuitionistic and
classical linear logic~\cite{cairespfenning,wadler2014propositions}.
Furthermore, linearly-typed functional languages then provide an excellent basis
for session-typed programming~\cite{DBLP:journals/jfp/GayV10}.

Modern substructural type systems, however, allow
us to go beyond the notion of linearity, classifying usage into more than just
linear or non-linear. This idea originates from Bounded Linear Logic (BLL),
which generalises the $!$ modality to a family of modalities indexed by a
polynomial expressing an upper bound on usage~\cite{girard1992bounded}. For
example, $!_{x \leq 2} A$ describes a proposition $A$ which can be used at most
twice.

From various directions this notion of BLL has been generalised further to
develop \emph{coeffect types}~\cite{brunel2014coeffects, ghica2014coeffects,
petricek2014coeffects}, which are a framework for capturing different kinds of
resource analysis in a single type system. \emph{Graded modal
types}~\cite{orchard2019quantitative} unify coeffects with their dual notion of
\emph{effects}, and are an expressive system allowing for the specification and
verification of many behavioural properties of programs. In this paper, we
demonstrate that combining session types with graded modal types allows for
various non-linear behaviours of concurrent programs to be reintroduced in a
linear setting, in a controlled and precise way.

A key part of this work is to reconcile the tension between three competing
requirements: (1) side effects inherent in communication primitives, (2)
non-linearity, and (3) the call-by-value semantics one might expect in most
programming languages (in contrast to call-by-name, which is the basis of most
theoretical explanations of linear and graded type systems). Requirements (1)
and (2) can be satisfied more easily in a call-by-name setting, but (3) (CBV
semantics) provides unsoundness. This is discussed in more detail in
Section~\ref{subsec:cbv}, where we also describe our chosen solution. A previous
iteration of session types for Granule was described by Orchard et
al.~\cite{orchard2019quantitative}. However, the system described required a
monadic interface to avoid some issues caused by CBV; we solve these problems
here to provide a more general and powerful interface for session-typed
programming, and also demonstrate the broad range of possible applications for
this interface by introducing a suite of primitives which capture the non-linear
behaviours of reuse, replication, and repetition (providing \emph{multicasting}).

Related ideas appear in the type system of Zalakain and Dardha
who use leftover typing to define a resource-aware
session type calculus that can represent shared, graded and linear
channels~\cite{zalakain}. We capture some of the same ideas via a
unified graded approach, showing how different amounts of sharing can
be characterised precisely with the interaction of linear, indexed,
and graded types.

\section{A Brief Granule and Graded Modal Types Primer}
\label{sec:primer}

Granule's type system is based on the linear $\lambda$-calculus augmented with
\emph{graded modal types}~\cite{orchard2019quantitative}. With linear typing as
the basis, we cannot write functions that discard or duplicate their inputs as
in a standard functional programming language. However, we can introduce
non-linearity via graded modalities and use these to represent such functions,
exemplified by the following Granule code:

\begin{minipage}{0.5\linewidth}
\begin{granule}
drop : forall {a : Type} . a [0] -> ()
drop [x] = ()
\end{granule}
\end{minipage}
\begin{minipage}{0.5\linewidth}
\begin{granule}
copy : forall {a : Type} . a [2] -> (a, a)
copy [x] = (x, x)
\end{granule}
\end{minipage}
The function arrow can be read as the type of linear functions (which consume
their input exactly once), but \granin{a [r]} is a graded modal type capturing
the capability to use the value `inside' in a non-linear way as described by
\granin{r}, i.e., \granin{drop} uses the value $0$ times and \granin{copy} uses
it $2$ times. The pattern match \granin{[x]} on the left-hand side of each
function eliminates the graded modality, binding \granin{x} as a non-linear
variable.

The central idea of graded modal types is to capture program structure
via an indexed family of modalities where the indices have some
algebraic structure which gives an abstract view of program
structure. We focus on \emph{semiring graded necessity} in this paper
(written $\Box_r A$ in mathematical notation but \granin{A [r]} in
Granule) which generalises linear logic's $\oc$~\cite{girard}
and Bounded Linear Logic~\cite{girard1992bounded}.
The above example uses the natural number graded modality,
counting exactly how many times a value can be used.

Another useful graded modality has grades drawn from a semiring of
\emph{intervals} which allows us to give bounds on how a value might be used. To
demonstrate, we define the classic \granin{fromMaybe} function which allows for
retrieving a value that may or may not exist from a \granin{Maybe} type.
\begin{granule}
data Maybe a = Just a | Nothing

fromMaybe : forall {a : Type} . a [0..1] -> Maybe a -> a
fromMaybe [_] (Just x) = x;
fromMaybe [d] Nothing  = d
\end{granule}
Note that without the graded modality, this function would be ill-typed in
Granule, since values are linear by default and one of the cases requires
discarding the default value (given by the first parameter).
By giving this parameter a type of \granin{a
[0..1]}, we specify that it can be used \emph{either} \granin{0} times or
\granin{1} time (in other words, this value has \emph{affine} behaviour rather
than linear). There is only one total function in Granule which inhabits the
type we give to \granin{fromMaybe} here--linearity forbids defining an instance
which always returns the default value.

In order to use \granin{fromMaybe} we need to `promote' its first
input to be a graded modal value, which is written by wrapping a value
in brackets, e.g., \granin{fromMaybe [42]}. Promotion propagates any
requirements implied by the graded modality to the free variables.
For example, the following takes an input and shares the capabilities
implied by its grade to two uses of \granin{fromMaybe}:
\begin{granule}
fromMaybeIntPlus : Int [0..2] -> Maybe Int -> Maybe Int -> Int
fromMaybeIntPlus [d] x y = fromMaybe [d] x + fromMaybe [d] y
\end{granule}
The \granin{0..1} usage implied by the first \granin{fromMaybe [d]}
and \granin{0..1} usage by the second are added together to get the
requirement that the incoming integer \granin{d} is graded as \granin{0..2}.

Lastly, Granule includes \emph{indexed} types which offer a lightweight form of
dependency, allowing for type-level access to information about
data. For example length-indexed vectors 
can be defined and used:

\begin{granule}
data Vec (n : Nat) (a : Type) where
  Nil  : Vec 0 a;
  Cons : a -> Vec n a -> Vec (n + 1) a

append : forall {a : Type, n m : Nat} . Vec n a -> Vec m a -> Vec (n + m) a
append Nil ys         = ys;
append (Cons x xs) ys = Cons x (append xs ys)
\end{granule}
Indexed types allow us to ensure at the type-level that when we append two
vectors, the length of the output is equal to the sum of the lengths of the two
inputs. Again, thanks to linearity, we gain further assurances from our type
signatures--here we can guarantee that since we must use every element of the
input vectors, these must also appear in the output, and so no information
is being lost along the way.

\paragraph{Type system redux}

The core type theory (based on Orchard et al.~\cite{orchard2019quantitative})
 extends the linear $\lambda$-calculus with
the semiring-graded necessity modality $\Box_r A$. Typing contexts
contain both linear assumptions $x : A$ and graded assumptions $x : [A]_r$
which have originated from inside a graded modality.
The core typing rules for introducing and eliminating graded modal
types are then as follows:
\begin{align*}
  \dfrac{\Gamma \vdash t : A}
  {r * \Gamma \vdash [t] : \Box_r A}
  \textsc{pr}\;\;
  \dfrac{\Gamma \vdash t_1 : \Box_r A \qquad \Delta, x : [A]_r \vdash t_2 : B}
  {\Gamma + \Delta \vdash \textbf{let}\ x = t_1\ \textbf{in}\ t_2 : B}
  \textsc{elim}\;\;
  \dfrac{\Gamma, x : A \vdash t : B}
  {\Gamma, x : [A]_1 \vdash t : B}
  \textsc{der}
  \;\;
  \dfrac{\Gamma \vdash t : B}
        {\Gamma, 0*\Delta \vdash t : B}
  \textsc{weak}
\end{align*}
The \textsc{pr} rule (promotion) introduces
a graded modality with grade $r$, and thus must scale by
$r$ all of the inputs (none of which are allowed to be linear as
$r * \Gamma$ is partial: it scales each graded assumption in $\Gamma$ by $r$ or
it is undefined if $\Gamma$ contains linear assumptions).
The \textsc{elim} rule captures the idea that a requirement for
$x$ to be used in an $r$-like way in $t_2$ can be matched
with the capability of $t_1$ as given by the graded modal type.
In Granule, this construct
is folded into pattern matching (seen above);
we can `unpack' (eliminate) a graded modality via pattern matching to provide
a non-linear (graded) variable in the body of the function
(the analogue to $t_2$ in this rule). The \textsc{der} rule connects linear typing
to graded typing, showing that a requirement for a linear
assumption is satisfied by an assumption graded by $1$.
Lastly, \textsc{weak} explains how we can \emph{weaken} with variables
graded by $0$.

Implicit in any rule involving multiple terms
is a notion of \emph{contraction} captured by the $+$ operation
on contexts, which is only defined when contexts are disjoint
with respect to linear assumptions,
and on overlapping graded assumptions we add their grades, e.g. $(\Gamma, x :
[A]_r) + (\Delta, x : [A]_s) = (\Gamma + \Delta), x : [A]_{r + s}$.

\section{Linear Session Types in Granule}
\label{sec:core}

\newcommand{\chan}[1]{\mathsf{Chan}(#1)}

The implementation of session types in Granule is based on the GV calculus
(originating from Gay and Vasconcelos~\cite{DBLP:journals/jfp/GayV10}, further
developed by Wadler~\cite{wadler2014propositions}, for which we use Lindley and
Morris' formulation~\cite{lindley2015semantics}). The GV system extends the linear
$\lambda$-calculus with a data type of channels $\chan{S}$ parameterised by session
types $S$~\cite{yoshida2007language}, which capture the protocol of interaction
allowed over the channel. Communication is asynchronous (sending always succeeds,
but receiving can block as usual).

Starting with a simple subset, session types can describe channels which send or
receive a value of type $T$ and then can be used according to session type $S$
written $!T . S$ or $?T . S$ respectively, or can be closed written
$\mathsf{end}_!$ or $\mathsf{end}_?$. There are then functions for sending or
receiving a value on a channel, forking a process at one end of a channel
returning the other, and waiting for a channel to be closed:
\begin{align*}
\begin{array}{llll}
\mathsf{send} & : \ T \otimes \chan{!T . S} \multimap \chan{S} \qquad
& \mathsf{fork} & : \ (\chan{S} \multimap \chan{\mathsf{end}_!}) \multimap \chan{\overline{S}} \\
\mathsf{recv} & : \ \chan{?T . S} \multimap T \otimes \chan{S} &
\mathsf{wait} & : \ (\chan{\mathsf{end}_?} \multimap 1)
\end{array}
\end{align*}
where $\overline{S}$ is the \emph{dual} session type to $S$, defined
$\overline{!T.S} = ?T.\overline{S}$, $\overline{?T.S} = !T.\overline{S}$,
$\overline{\mathsf{end}_?} = \mathsf{end}_!$ and $\overline{\mathsf{end}_!} =
\mathsf{end}_?$. The $\mathsf{fork}$ combinator leverages the duality operation,
spawning a process which applies the parameter function with a fresh channel,
thus returning the dual endpoint.

The core interface in Granule correspondingly has operations with the
following types, where \granin{Protocol} is the kind of session types
whose constructors we highlight in \textcolor{purple!70!black}{purple}:
\begin{granuleInterface}
send        : forall {a : Type, p : Protocol} . LChan (Send a p) -> a -> LChan p
recv        : forall {a : Type, p : Protocol} . LChan (Recv a p) -> (a, LChan p)
forkLinear  : forall {p : Protocol}           . (LChan p -> ())  -> LChan (Dual p)
close       : LChan End -> ()
\end{granuleInterface}
We also provide some utility operations for internal and external choice (also
known as selection and offering); note that as described in the original
formulation by Lindley and Morris~\cite{lindley2015semantics} these can be
defined in terms of the core functions given above, but we provide primitives
for ease of use.
\begin{granuleInterface}
selectLeft   : forall {p1 p2 : Protocol} . LChan (Select p1 p2) -> LChan p1
selectRight  : forall {p1 p2 : Protocol} . LChan (Select p1 p2) -> LChan p2
offer        : forall {p1 p2 : Protocol, a : Type}
             . (LChan p1 -> a) -> (LChan p2 -> a) -> LChan (Offer p1 p2) -> a
\end{granuleInterface}
The following gives a brief example putting all these primitives
together:
\begin{granule}
server : LChan (Offer (Recv Int End) (Recv () End)) -> Int [0..1] -> Int
server c [d] = offer (\c -> let (x, c) = recv c in let () = close c in x)
                     (\c -> let ((), c) = recv c in let () = close c in d) c

client : forall {p : Protocol} . LChan (Select (Send Int End) p) -> ()
client c = let c = selectLeft c;
               c = send c 42
           in close c

example : Int -- Evaluates to 42
example = server (forkLinear client) [100]
\end{granule}
The \granin{server} offers a choice between being sent an integer or a unit
value. The second parameter (bound to \granin{d}) is used as a default value in
the case that a unit value is received by the server, where \granin{Int [0..1]}
denotes that this integer can can be used 0 or 1 times (see
Section~\ref{sec:primer} for more explanation of this grading). The
\granin{client} selects the left behaviour, sends an integer, then closes its
side of the communication.

The final definition \granin{example} spawns
the client with a channel and connects the dual end of the channel to
the server which returns the received value of \granin{42} here.

\section{The Relationship Between Grading, Call-by-Value, and Effects}
\label{subsec:cbv}

Consider the following example which is allowed by the type system
described so far:
\begin{granule}
problematic : Int
problematic =
  let [c] : ((LChan (Recv Int End)) [2]) = [forkLinear (\c -> close (send c 42))];
      (n, c') = recv c;  () = close c';
      (m, c') = recv c;  () = close c'
   in (n + m)
\end{granule}
On line 3, the program forks a process that sends $42$ on a channel, but under a promotion, with
the type explaining that we want to use the resulting value twice
(given by the explicit type signature here). This promotion then
allows two uses of the channel on lines 4-5.

The typical semantics for coeffect-based calculi in the literature is
call-by-name~\cite{brunel2014coeffects, gaboardi2016combining}. Under a
call-by-name semantics, which Granule allows via the extension \granin{language
CBN}, this program executes and produces the expected result of \granin{84}. The
key is that call-by-name reduction substitutes the call to \granin{forkLinear}
into the two uses of variable \granin{c} on lines 4 and 5, and thus we are
receiving from two different channels. However, under the default call-by-value
semantics (which was chosen in Granule for simplicity and to avoid complications
resulting from effects), line 3 is fully evaluated (underneath the graded modal
introduction), and so variable \granin{c} on lines 4 and 5 refers to a single
channel. This means that executing this program blocks indefinitely on line 5;
we get an error from the underlying implementation's concurrency
primitives (written in Haskell) describing a
\texttt{thread blocked indefinitely in an MVar operation}. This comes from
Granule's runtime which leverages these standard primitives.

In order to get around this problem with non-linear channels in a call-by-value
setting, Orchard et al.~\cite{orchard2019quantitative} instead deal with
channels monadically, so here we would end up with a channel of type
\granin{(Session (LChan ...)) [2]} (with the \granin{Session}
monad).\footnote{In Granule such a channel would in fact have the type
\granin{(LChan ... <Session>) [2]}, with \granin{<Session>} being an instance of
the graded monadic \granin{<r>} modalities which are dual to the graded comonadic
\granin{[r]} modalities. We elide discussing these further here since we will not
need any monad other than \granin{Session}.} To make this example well-defined
would then require a distributive law which maps from this to \granin{Session
((LChan ... ) [2])}, copying the channel. Providing such a distributive law is
unsound---it would enable \granin{problematic} and thus indefinite
blocking---and fortunately it is also not derivable. However, we wish to avoid
the monadic programming style as it is not required when working with just
linearity.

Our alternative solution is that we instead simply syntactically
restrict promotion for primitives which create linear channels; in particular
this includes the \granin{forkLinear} function which causes the difficulty here.
Non-linear channels are then re-introduced by additional primitives in
Section~\ref{sec:non-linear-patterns} which allow for precise and carefully
managed non-linear usage, allowing for grading to be combined with linear
channels even in a call-by-value setting and without the extra overhead that was
required when every channel had to be wrapped in the \granin{Session} monad.

\section{Non-linear Communication Patterns via Grading}
\label{sec:non-linear-patterns}

We show three common patterns and how they can be described by
combining linear session types and graded modal types:
\emph{reusable channels} (Section~\ref{subsec:reuse}),
\emph{replicated servers} (Section~\ref{subsec:replicated}),
and
\emph{multicast communication} (Section~\ref{subsec:multicast}).
All of these patterns are a variation
on the \granin{forkLinear} primitive, with some amount of substructural
behaviour introduced via graded modal types, and sometimes restrictions
to protocols via \emph{type predicates}. In Granule these are represented
similarly to Haskell's type class constraints with type signatures of
the form: \granin{functionName : forall \{typeVariables\} . \{constraints\} => type}.

\subsection{Reusable Channels}
\label{subsec:reuse}

A reusable or \emph{non-linear} channel is one which can be shared and thus
used repeatedly in a sound way. This contrasts with the notion seen in
Section~\ref{subsec:cbv} of promoting a fresh linear channel to being non-linear
in a call-by-value setting, which was unsound; a linear channel used in a shared
way can easily lead to a deadlock where one user of the channel leaves it in a
state which is then blocking for another user of the channel. The key issue with
sharing a channel is ending up with inconsistent states across different shared
usages. This can be avoided if the protocol allowed on the channel is restricted
such that only a single `action' (send, receive, choice, or offer) is allowed,
and thus if the channel is used multiple times it can never be left in an
inconsistent state across shared uses. This is captured by the idea that $(P \mid
P) \neq P$ in general in process calculi and so a replicated channel
which has more than a single action cannot be split off into many parallel uses,
e.g., $(ab)^* \mid (ab)^* \not\equiv (ab)^*$. However, if there is only a single
action then multiple repeated parallel uses are consistent, e.g., $a^* \mid a^*
\equiv a^*$.

We capture this idea via the following variant of the fork primitive with
arbitrarily graded channels:
\begin{granuleInterface}
forkNonLinear : forall {p : Protocol, s : Semiring, r : s} . {SingleAction p}
             => ((LChan p) [r] -> ()) -> (LChan (Dual p)) [r]
\end{granuleInterface}
where \granin{SingleAction : Protocol -> Predicate} is a type
constraint that characterises only those
protocols which comprise a single send, receive, choice, or offer,
i.e.,
\begin{align*}
\setlength{\arraycolsep}{1.5em}
  \begin{array}{lll}
  \text{\granin{SingleAction End}}
  &
  \text{\granin{SingleAction (Send a End)}}
  &
  \text{\granin{SingleAction (Offer End End)}} \\[-0.25em]
  & \text{\granin{SingleAction (Recv a End)}}
  & \text{\granin{SingleAction (Select End End)}}
  \end{array}
\end{align*}
As an example, consider a channel in a graded modality which
says it can be used exactly \granin{n} times. The following
uses this channel to send every element of a vector of size \granin{n}
(using the \granin{Vec} type of Section~\ref{sec:primer}):
\begin{granule}
sendVec : forall {n : Nat, a : Type} . (LChan (Send a End)) [n] -> Vec n a -> ()
sendVec [c] Nil         = ();
sendVec [c] (Cons x xs) = let () = close (send c x) in sendVec [c] xs
\end{granule}
This code shows the powerful interaction between grading, linearity, and indexed
types in Granule. Note that the above code is typeable without any of the new
primitives described in this section, but would be unusable without promoting a
use of \granin{forkLinear}. However, we can complete this example with a dual
process (\granin{recvVec}) that is then connected to \granin{sendVec} via
\granin{forkNonLinear}:
\begin{granule}
recvVec : forall {n : Nat, a : Type} . N n -> (LChan (Recv a End)) [n] -> Vec n a
recvVec Z [c]     = Nil;
recvVec (S n) [c] = let (x, c') = recv c; () = close c' in Cons x (recvVec n [c])

example : forall {n : Nat, a : Type} . Vec n a -> Vec n a
example xs = let (n, list) = length' list
             in recvVec n (forkNonLinear (\c -> sendVec c list))

main : Vec 5 Int
main = example (Cons 1 (Cons 1 (Cons 2 (Cons 3 (Cons 5 Nil)))))
\end{granule}
Note that the receiver needs to know how many elements to receive, so
this information has to be passed separately (via an indexed natural
number \haskin{N n}). A system with dependent
session types~\cite{DBLP:conf/ppdp/ToninhoCP21,DBLP:conf/fossacs/ToninhoY18}
could avoid this by first sending the length, but this is not (yet)
possible in Granule; the Gerty prototype language provides full dependent types
and graded modal types which would be a good starting
point~\cite{DBLP:conf/esop/MoonEO21}.

\subsection{Replicated Servers}
\label{subsec:replicated}

The $\pi$-calculus provides replication of a process $P$ as the process $!P$,
which session-typed $\pi$-calculus variants have refined into the more
controlled idea of having a replicated `server'. Here, the spawning of
replicated instances is controlled via a special
receive~\cite{berger2001sequentiality,sangiorgi2003pi}, e.g., written like
$*c(x).P$ meaning receive an $x$ on channel $c$ and then continue as $P$, whilst
still providing the original process. From an operational semantics point of
view this looks like $*c(x).P \mid \overline{c}\langle{d}\rangle.Q \rightarrow
*c(x).P \mid P[d/x] \mid Q$ where $\overline{c}\langle{d}\rangle.Q$ sends the
message $d$ to the server which `spawns' off a fresh copy of the server process
$P$ whilst also preserving the original server process for further clients to
interact with.

We provide this functionality here as the fork variant
\granin{forkReplicate}:
\begin{granuleInterface}
  forkReplicate : forall {p : Protocol, n : Nat} . {ReceivePrefix p}
               => (LChan p -> ()) [0..n] -> N n -> Vec n ((LChan (Dual p)) [0..1])
\end{granuleInterface}
Here the grading is less general than in Section~\ref{subsec:reuse};
we instead focus on a particular grading which says that given
a server process \granin{(LChan p -> ())} that can be used $0$ to $n$ times,
then we get a vector of size $n$ of dual channels which we can use
to interact with the server. The predicate \granin{ReceivePrefix p}
classifies those protocols which start with a receive, which includes
both \granin{ReceivePrefix (Recv a p)} and \granin{ReceivePrefix (Offer p1 p2)}.

Each client channel can itself be discarded due
to the graded modality \granin{... [0..1]}. Thus, we can choose not to use
any/all of the client channels, reflected in the dual side where the
server can be used at most $n$ times. This introduces some flexibility
in the amount of usage. A more strict variant is given as:
\begin{granuleInterface}
  forkReplicateExactly : forall {p : Protocol, n : Nat} . {ReceivePrefix p}
                      => (LChan p -> ()) [n] -> N n -> Vec n (LChan (Dual p))
\end{granuleInterface}
meaning we have exactly $n$ clients that \emph{must} spawn the server
$n$ times.

The following example demonstrates \granin{forkReplicate}
in action with two clients.

\begin{granule}
addServer : LChan (Offer (Recv Int (Recv Int (Send Int End))) (Recv Int (Send Bool End))) -> ()
addServer c =
 offer (\c -> let (x, c) = recv c;  (y, c) = recv c; in close (send c (x + y))
       (\c -> let (x, c) = recv c;                   in close (send c (x == 0))) c

client1 : forall {p : Protocol} . LChan (Select (Send Int (Send Int (Recv Int End))) p) -> Int
client1 c = let (x, c) = recv (send (send (selectLeft c) 10) 20); () = close c in x

client2 : forall {p : Protocol} . LChan (Select p (Send Int (Recv Bool End))) -> Bool
client2 c = let (b, c) = recv (send (selectRight c) 42);  () = close c in b

import Parallel -- Provides a `par` combinator derived from `forkLinear`
main : (Int, Bool)
main = let (Cons [c1] (Cons [c2] Nil)) = forkReplicate [addServer] (S (S Z))
       in par (\() -> client1 c1) (\() -> client2 c2)
\end{granule}
Here \haskin{addServer} provides functionality offering the
ability to receive two integers and send back their addition,
or receive a single integer and send back whether it is a zero
or not. The two clients \haskin{client1} and \haskin{client2}
exercise both behaviours, via channels given
by \granin{forkReplicate} of \haskin{addServer}, and are run in parallel on line 15
using the \granin{par} combinator which itself implemented in
terms of \granin{forkLinear}.

In the intuitionistic linear logical propositions of Caires and
Pfenning~\cite{cairespfenning}, this same idea is captured by a non-linear
channel (in Dual Intuitionistic Linear Logical style) which yields a linear
version when interacted with; this is akin to the idea of \emph{shared
channels}~\cite{yoshida2007language}. Our approach codifies the same principle
using graded modalities, but only allows finite replication
(Section~\ref{sec:discussion} discusses unbounded replication).

\subsection{Multicast Sending}
\label{subsec:multicast}

The final primitive we introduce here provides the notion of \emph{multicast} (or
\emph{broadcast}) communication where messages on a channel can
be received by multiple clients. The non-linearity here is now
on the payload values being sent, with grading to explain that the amount of non-linearity matches the number
of clients. The \granin{forkMulticast} primitive provides this
behaviour:
\begin{granuleInterface}
forkMulticast : forall {p : Protocol, n : Nat}
              . {Sends p} => (Chan (Graded n p) -> ()) -> N n -> Vec n (Chan (Dual p))
\end{granuleInterface}
where \granin{Graded : Nat -> Protocol -> Protocol} is a type function
adding a graded modality to payload types:
\begin{align*}
\setlength{\arraycolsep}{0.2em}
\begin{array}{ll}
  \begin{array}{ll}
\text{\granin{Graded n (Send a p)}} & = \ \text{\granin{Send (a [n]) (Graded n p)}} \\[-0.25em]
\text{\granin{Graded n (Recv a p)}} & = \ \text{\granin{Recv (a [n]) (Graded n p)}}\\[-0.25em]
\text{\granin{Graded n (Select p1 p2)}} & = \ \text{\granin{Select (Graded n p1) (Graded n p2)}}\\[-0.25em]
\text{\granin{Graded n (Offer p1 p2)}} & = \ \text{\granin{Offer (Graded n p1) (Graded n p2)}}
\end{array} &
\begin{array}{ll}
  \text{\granin{Graded n End}} & = \ \text{\granin{End}} \\ \\ \\
\end{array}
\end{array}
\end{align*}
and \granin{Sends : Protocol -> Predicate} is defined:
\begin{align*}
\dfrac{\quad}{\text{\granin{Sends End}}}
\qquad
\dfrac{\text{\granin{Sends p}}}
      {\text{\granin{Sends (Send a p)}}}
\qquad
\dfrac{\text{\granin{Sends p1}} \qquad \text{\granin{Sends p2}}}
      {\text{\granin{Sends (Select p1 p2)}}}
\end{align*}
Thus, as long as we are sending values that are wrapped in the graded
modality such that they can be used $n$ times, we can then broadcast
these to $n$ client participants via the \granin{Vec n (Chan (Dual
  p))} channels.

For example, in the following we have a \granin{broadcaster} that takes a
channel which expects an integer to be sent which can be used 3 times.
\begin{granule}
broadcaster : LChan (Send (Int [3]) End) -> ()
broadcaster c = close (send c [42])
\end{granule}
We can then broadcast these results
with \granin{forkMulticast broadcaster (S (S (S Z)))} producing
 a 3-vector of channels. Below we aggregate the results from these
three receiver channels by applying \granin{aggregateRecv}
to the vector, giving us the result \granin{126}, i.e. \granin{3 * 42}.
\begin{granule}
aggregateRecv : forall {n : Nat} . Vec n (LChan (Recv Int End)) -> Int
aggregateRecv Nil         = 0;
aggregateRecv (Cons c cs) = let (x, c) = recv c; () = close c in x + aggregateRecv cs

main : Int -- Evaluates to 126
main = aggregateRecv (forkMulticast broadcaster (S (S (S Z))))
\end{granule}
\section{Discussion and Conclusion}
\label{sec:discussion}

The key idea here is that graded types capture various standard non-linear forms
of communication pattern atop the usual linear session types presentation. This
differs, but is related to, the generalisation of session types via adjoint
logic presented by Pruiksma and Pfenning~\cite{pruiksma}, which offers
non-linearity but without the precise quantification allowed by our grades.
The demands of call-by-value required a different approach to past work, with
syntactic restriction discussed in Section~\ref{subsec:cbv} and vectors to capture
multiplicity of channels. The next step would be to prove type safety given the
additional restriction on promoting channels and the other features introduced
here.

\paragraph{Recursion and other combinators}

A notable omission from our core session types calculus is the ability
to define \emph{recursive protocols}. However, some of the power of
recursive session types is provided here;
reusable channels (Section~\ref{subsec:reuse}) are equivalent to
linearly recursive session types (e.g., $\mu x . P.x$), especially when
the grade \granin{r} is instantiated to \granin{0..Inf} to capture an
arbitrary amount of use. Further work is to integrate standard ideas
on recursive session types and to explore their interaction with grading.

Combinations of the ideas here are also possible; for example, combining
multicast sending with reuse to get channels which we can repeatedly use
to broadcast upon. We could also define a more powerful version of
replication to allow an unknown number of clients (possibly infinite),
with an interface similar to \granin{forkReplicate} but returning a lazy stream
of client channels which can be affinely used:
\begin{granuleInterface}
  forkReplicateForever : forall {p : Protocol} . {ReceivePrefix p}
                      => (LChan p -> ()) [0..Inf] -> Stream ((LChan (Dual p)) [0..1])
\end{granuleInterface}
\paragraph{Further applications and related ideas}

Using the linear channels already present in Granule it is possible to represent
functions that are \emph{sequentially realizable}~\cite{inverses}; a
sequentially realizable function is one which has outwardly pure behaviour but
relies on a notion of local side effects which are contained within the body of
the function. Looking into which such behaviours may be more easily expressed by
introducing non-linearity via graded channels would be an avenue for future
work.

As discussed in Section~\ref{subsec:cbv}, it is necessary to restrict promotion
of channels due to Granule's default call-by-value semantics, since otherwise
this allows for a linear channel to be used non-linearly. A very similar caveat
applies if we consider mutable arrays. Recent work describes how to represent
uniqueness of reference within Granule in addition to linearity~\cite{entente};
if a unique array is promoted then we end up with multiple references to the
same array, so we can no longer guarantee uniqueness for mutation. This is
resolved by restricting promotion in much the same way as we have here for
linear channels.

Uniqueness in a concurrent setting was considered in various works by de Vries
et al., but in particular we draw attention to their paper on resource
management via unique and affine channels~\cite{devries}. They allow for a
notion of a channel which we can guarantee unique access to after $i$
communication steps where $i$ is some natural number; this is a form
of grading, though quite different from the ideas we have discussed. It would be
interesting to explore how we can represent this in Granule which already has
both uniqueness and graded session types, and whether we gain further
expressivity from doing so.

We explored combining grading with linear session types in Granule, but this
could be extended to other settings. The linear types extension to Haskell is
based on a calculus involving graded function arrows~\cite{linearhaskell}, and
adding additional multiplicities to this system is a possibility. The Priority
Sesh library provides a convenient embedding of the GV linear session calculus
in Haskell~\cite{prioritysesh}, and extending this to make use of more precise
information about channel usage could be valuable. This would also allow for
experimenting with graded channels in a setting where grading can
be implicit, rather than one where (like Granule) all grades are explicitly
encoded via modalities.

\paragraph{Acknowledgments}
This work was supported by an EPSRC Doctoral Training Award (Marshall) and EPSRC
grant EP/T013516/1 (\emph{Verifying Resource-like Data Use in Programs via
Types}).

\nocite{*}
\bibliographystyle{eptcs}
\bibliography{references}
\end{document}